\documentclass[aps,prb,twocolumn,eqsecnum,superscriptaddress,floatfix,10pt]{revtex4-2}
\usepackage{amsmath,amssymb,amsfonts,stmaryrd,wasysym,graphicx,multirow,color,textcomp,subfigure,nicefrac,float,enumitem,physics}

\usepackage{url}%\usepackage{diagrams}
\usepackage[colorlinks=true,citecolor=blue,urlcolor=blue,linkcolor=blue]{hyperref}

\def\be{\begin{equation}}
\def\ee{\end{equation}}
\def\bea{\begin{eqnarray}}
\def\eea{\end{eqnarray}}

\counterwithout{equation}{section}

\begin{document}

\title{$\eta$-pairing state in flatband lattice: \\Interband coupling effect on entanglement entropy logarithm}
\author{Seik Pak}
\address{Department of Physics, Hanyang University, Seoul 04763, Republic of Korea}

\author{Hanbyul Kim}
\address{Department of Physics, Hanyang University, Seoul 04763, Republic of Korea}

\author{Chan Bin Bark}
\address{Department of Physics, Hanyang University, Seoul 04763, Republic of Korea}

\author{Sang-Jin Sin}
\email{sjsin@hanyang.ac.kr}
\address{Department of Physics, Hanyang University, 222 Wangsimni-ro, Seoul, 04763, Public of Korea}

\author{Jae-yoon Choi}
\email{jaeyoon.choi@kaist.ac.kr}
\address{Department of Physics, Korea Advanced Institute of Science and Technology, Daejeon 34141, Korea}

\author{Moon Jip Park}
\email{moonjippark@hanyang.ac.kr}
\address{Department of Physics, Hanyang University, Seoul 04763, Republic of Korea}
\address{Research Institute for Natural Science and High Pressure, Hanyang University, Seoul, 04763, South Korea}

\date{\today}

\begin{abstract}
The $\eta$-pairing state is the eigenstate of the hypercubic Hubbard model, which exhibits anomalous logarithmic scaling of entanglement entropy. In multi-band systems, $\eta$-pairing can be exact eigenstate when the band is flat without interband coupling. However, typical flatband systems such as Lieb and Kagome lattices often feature band touchings, where interband coupling effects are non-negligible. Using the Creutz ladder, we investigate the deformation of $\eta$-pairing states under the interband coupling effect. Our results show corrections to entanglement entropy scaling, with modified $\eta$-pairing states displaying broadened doublons, nonuniform energy spacing, and deviations from exact behavior for configurations with more than one $\eta$-pair, even in the large band gap limit, except at $t = 0$. Through a Schrieffer--Wolff transformation, we quantify corrections to the spectrum generating algebra, offering insights into the interplay between interaction-driven phenomena and band structure effects. These findings illuminate the robustness and limitations of $\eta$-pairing in realistic flatband systems.

\end{abstract}

\maketitle

\section{Introduction}

Entanglement entropy (EE) is a fundamental measure of quantum correlations in diagnosing many-body quantum phases~\cite{RevModPhys.82.277,EE1,EE2,scarEE,EE.98.220603,EE121,EE123,etaee.6.023041}.  In general, EE of thermal states obeys a volume law, scaling proportionally to the system size ($\sim L^d$), while the ground states of gapped Hamiltonians typically follow an area law ($\sim L^{d-1}$). On the other hand, certain critical systems~\cite{PhysRevLett.97.050404,PhysRevLett.93.260602,PhysRevLett.90.227902,critical.10.011047,critical.79.115421,critical.96.100603} exhibit logarithmic scaling of EE ($\sim \log L$), reflecting the presence of long-range correlations and emergent symmetry structures such as quantum many-body scars and Hilbert space fragmentation~\cite{Moudgalya_2022,Serbyn2021,PhysRevLett.122.173401,scar1,scar3,scar4,scar2,scarEE,HSF,HSF.10.011047,HSF.12.011050,HSF.124.207602}.

A representative example of such emergent structure is the $\eta$-pairing states, first introduced by C. N. Yang in the context of the single-band Hubbard model~\cite{PhysRevLett.63.2144}. A hidden pseudo-spin SU(2) symmetry ensures the perfectly bound Cooper pairs as the eigenstates giving rise to the off-diagonal long-range order (ODLRO) of a doublon pairing order. The resulting $\eta$-paired eigenstate form an equally spaced tower of eigenstates. In multi-band systems, recent work has shown that $\eta$-pairing can also be stabilized in flatband systems, when the flatband is perfectly isolated from other bands~\cite{torma.94.245149,tormaBernivig}.

In many lattice realizations of the flatbands such as Lieb, Kagome, and Creutz ladder models, the flatband accompanies symmetry protected band touching with dispersive bands~\cite{PhysRevB.78.125104,PhysRevB.99.045107}. This band touching induces non-negligible couplings between the flat and dispersive bands, thereby spoiling the exact pseudo-spin SU(2) symmetry that underpins the ideal $\eta$-pairing mechanism. Consequently, it remains an open question how $\eta$ pairing states and the associated dynamical phenomena, such as energy spacing, correlation length, and entanglement entropy, evolve under the interband interactions.

In this work, we address the modification of ideal $\eta$-pairing states in the Creutz ladder, a representative flatband model in one dimension. We reveal three key phenomena. First, even when the flatband is not strictly isolated, a distinct subset of eigenstates continues to exhibit log-law behaving entanglement entropy—indicative of $\eta$-like states. Second, while interband coupling disrupts the exact pseudo-spin SU(2) symmetry and its associated spectrum generating algebra (SGA) of the $\eta$-pairing states, Schrieffer–Wolff (SW) transformation demonstrates that an approximate SGA emerges, capturing the modified energy shifts induced by virtual excitations. Third, in real space, the originally on-site doublon pairs broaden into extended states as hybridization sets in, indicating that interband coupling effectively delocalizes the pair correlations.

Recent studies have further revealed that $\eta$-pairing states are intimately connected to quantum many-body scars (QMBS), a class of nonthermal eigenstates that coexist with a predominantly thermal spectrum~\cite{etascar1,PhysRevB.102.075132,etaee.6.023041,PhysRevResearch.7.013064,PhysRevResearch.3.043156,PhysRevLett.125.230602,PhysRevB.105.024520,PhysRevResearch.6.043259}. Importantly, interband coupling can bridge these otherwise isolated sectors, effectively modifying the interactions between macroscopic degrees of freedom and, consequently, the overall thermalization dynamics. Finally we discuss experimental realizations of our results in the cold atom systems.

\section{Model and Theoretical Setup} 
\label{Sec.Model and Theoretical Setup}

\subsection{$\eta$-pairing states in hypercubic lattice} 

We start our discussion by introducing the $\eta$-pairing state in the $D$-dimensional hyper-cubic lattice Hubbard model. The Hamiltonian is given as,
\bea
    H_{\textrm{cubic}} = t \sum_{\langle \textbf{r},\textbf{r}' \rangle, \sigma} c_{\textbf{r},\sigma}^\dagger c_{\textbf{r}',\sigma} + U\sum_\textbf{r}n_{\textbf{r},\uparrow} n_{\textbf{r},\downarrow}\notag  -\mu\sum_{\textbf{r},\sigma}c_{\textbf{r},\sigma}^\dagger c_{\textbf{r},\sigma},
\eea
where $c_{\mathbf{r},\sigma}$ ($c_{\mathbf{r},\sigma}^\dagger$) denotes the fermionic annihilation (creation) operator at site $\mathbf{r}$ and spin $\sigma$, with $n_{\mathbf{r},\sigma} = c_{\mathbf{r},\sigma}^\dagger c_{\mathbf{r},\sigma}$ representing the number operator. Here, $\langle \textbf{r},\textbf{r}' \rangle$ denotes nearest-neighbor pairs.

Assuming the periodic boundary conditions (PBC), we can define the $\eta$-operator, which forms a doublon pair with the center of mass momentum $\boldsymbol{\pi} = (\pi, \pi, \dots,\pi).$ as, \bea
    \eta = \sum_\mathbf{r} e^{i \boldsymbol{\pi} \cdot \textbf{r}} \, c_{\mathbf{r},\uparrow} c_{\mathbf{r},\downarrow}
    ,
\eea
The Hubbard model preserves pseudo-spin SU(2) symmetry. The generators of pseudo-spin SU(2) symmetry that are defined as, $  \eta_x = \frac{1}{2}(\eta+\eta^\dagger), \ \eta_y = \frac{i}{2}(\eta-\eta^\dagger), \ \eta_z = \frac{1}{2}[\eta^\dagger,\eta]$, satisfy the following symmetry algebra
\begin{align}
    [\eta_a,\eta_b] = i \epsilon_{abc} \eta_c,\ [H,J^2] = 0,\ [H,\eta_z] = 0
\end{align}
where $J^2 = \eta_x^2+\eta_y^2+\eta_z^2$ and $a,b$, and $c$ range over $x,y$, and $z$, respectively.
Accordingly, the $\eta$ operators satisfy the following commutation relation (also known as the SGA~\cite{PhysRevB.102.085140}).
\bea
\label{SGA}
[H,\eta^\dagger] = (U-2\mu)\eta^\dagger,
\eea
Due to the SGA, the eigenenergies of the Hubbard models forms  an equally spaced tower of eigenstates,
\bea
\label{tower}
| \psi_n \rangle = \bigl(\eta^\dagger\bigr)^n \,\ket{\psi_0},
\eea
where $\ket{\psi_0}$ is the vacuum (or a reference eigenstate) and $n$ runs over the possible number of pairs. Intuitively, each application of $\eta^\dagger$ creates a doublon pair, raising the energy by $U-2\mu$ due to the on-site interaction and chemical potential. This original formulation of $\eta$-pairing, characterized by a macroscopic doublon condensate with ODLRO, sets the stage for our investigation. 

$\eta$-pairing generates the integrable dynamics in the eigenstate spectrum, resulting in the logarithmic EE behavior. It can be predicted from quasiparticle nature of $\eta$-pairing states appeared in Eq.~\ref{tower} ~\cite{quasi}. From homogeneous spatial configuration of the doublons, one can write the reduced density matrix of the $\eta$-pairing states as follows,
\begin{align}
    \rho_A = \sum_{i = 0}^{L_A} \lambda_i \, |i\rangle\langle i|,\ \  \lambda_i = \frac{
    \begin{pmatrix}
        L_A \\ i
    \end{pmatrix} 
    \begin{pmatrix}
        L_B \\ n-i
    \end{pmatrix}}
    {\begin{pmatrix}
        L \\ n
    \end{pmatrix}},
\end{align}
where $L = L_A + L_B$ is the total system size, $L_A$ and $L_B$ denote the sizes of subsystems $A$ and $B$, respectively, and $\ket{i}$ represents a symmetric state with $i$ doublons in $L_A$ sites~\cite{Fan_2005,Vedral_2004}. The EE is then given by $S = -\sum_{i=0}^{L_A} \lambda_i \ln \lambda_i$. In the thermodynamic limit, this sum can be evaluated analytically, yielding
\begin{equation}
S = \frac{1}{2}\Bigl(1 + \ln\bigl[2\pi \tfrac{n}{L}\bigl(1 - \tfrac{n}{L}\bigr)L_A\bigr]\Bigr),
\end{equation}
which shows a characteristic logarithmic dependence on $L_A$~\cite{exactEE}.

\subsection{Multiband flatband systems} 

To capture the essential physics of $\eta$-pairing in flatband systems, we consider the Hamiltonian of the form $H_{\textrm{total}} = H_\mathrm{kin}-\mu N + H_\mathrm{int}$,
where $H_\mathrm{kin}$ denotes the single-particle tight-binding moel and $H_\mathrm{int}$ the on-site interaction term respectively. Without loss of generality, the tight-binding model can be written in the momentum space as, 
\bea
H_\mathrm{kin} = \sum_{n=1,s=\uparrow,\downarrow}^{N_{\textrm{orb}}} \sum_{\mathbf{k}\in\textrm{BZ}} \epsilon_{n}(\mathbf{k})\gamma^\dagger_{n,\mathbf{k},s}\,\gamma_{n,\mathbf{k},s},
\eea
where $\epsilon_{n,s}(\mathbf{k})$ is the eigenenergy of $n$-th band with momentum $\mathbf{k}$ and spin $s=\uparrow,\downarrow$. $\gamma_{n,\mathbf{k},s}$ is the annihilation operator of $n$-th single-particle eigenstates, where the electron operator is related with the eigenvector such that $c_{\alpha,\mathbf{k},s}=[U(\mathbf{k})]_{\alpha,n}\gamma_{n,\mathbf{k},s}$ with the orbital index $\alpha$.
\begin{figure}[t]
    \centering
    \includegraphics[width=\linewidth]{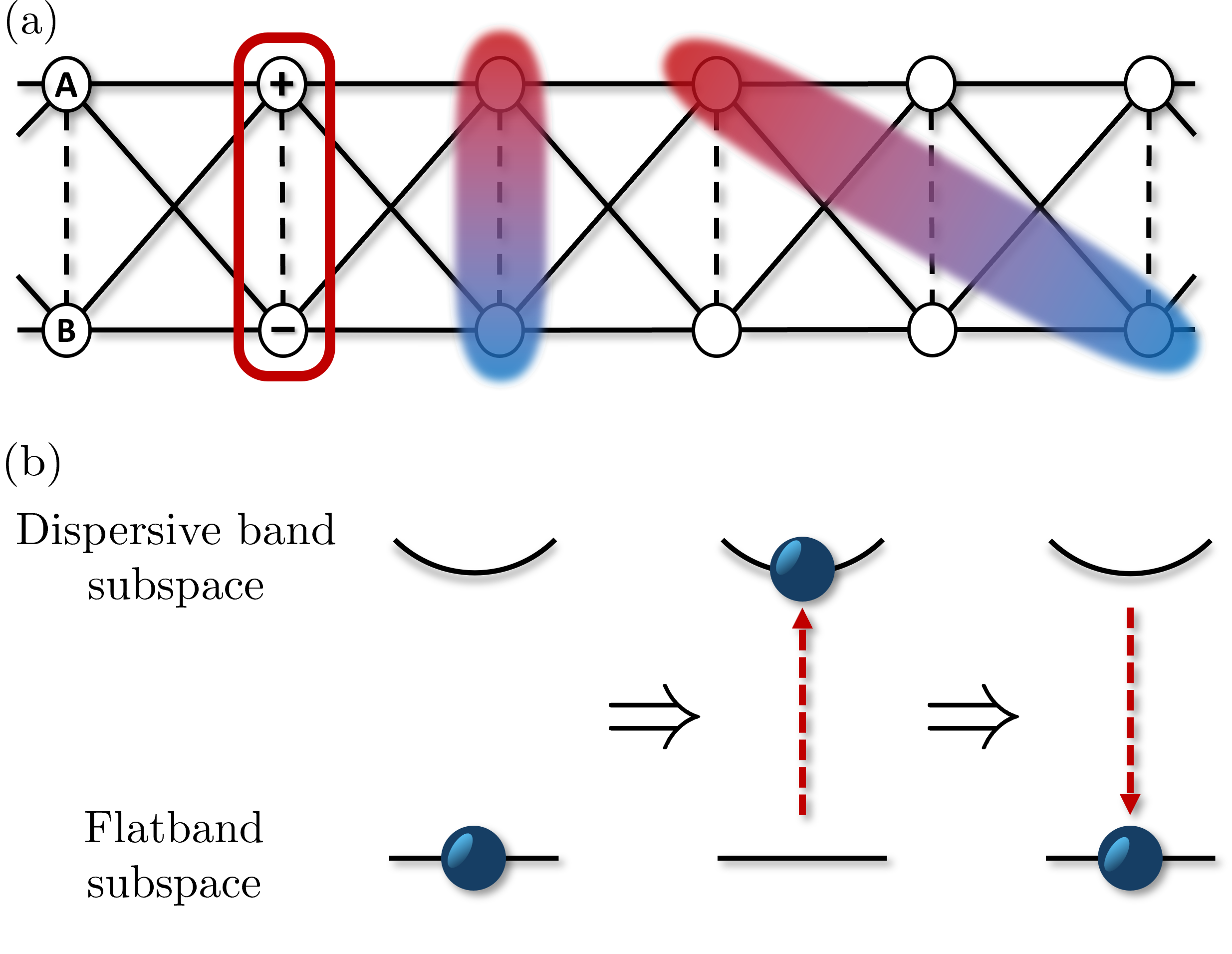}
    \caption{(a) Schematic representation of the Creutz ladder, consisting of two sublattices $A$ and $B$. Solid lines represents intercell and dashed lines denotes intracell hopping of amplitudes $t$ and $t'$, respectively. The red box highlights the CLS of the flatband. Shaded regions represent the spatial profile of doublon pairs: The vertical shaded region illustrates the localized doublon pair characteristic of the exact $\eta$-pairing state, while the tilted shaded region shows the spatially extended doublon pair associated with the modified $\eta$-pairing state, which arises due to interband coupling. (b) Schematic representation of virtual process in SW transformation between flatband and dispersive band subspace. Red arrows denote virtual process.}
    \label{fig:Creutz}
\end{figure}
Without loss of generality, we assume that $n=1$ is the flatband while the others are dispersive bands. We consider the proection of the full Hilbert space onto the flatband subspace. The projection operator is then defined as,
$
P_{\alpha \beta}(\mathbf{k}) = [U({\mathbf{k}})]_{\alpha,1} [U({\mathbf{k}})^\dagger]_{1,\beta},
$
and the Fourier transform yields the corresponding projected electron operator in real space:
\bea
\bar{c}_{i,\alpha,\sigma} 
&=&\frac{1}{\sqrt{N}} \sum_{\mathbf{k},\beta} e^{-i\mathbf{k}\cdot\mathbf{R}_i} P_{\alpha\beta}(\mathbf{k})c_{\mathbf{k},\beta,\sigma} \notag \\
&=&\sum_{j,\beta}P_{\alpha \beta}(i-j) c_{j,\beta,\sigma}\\
\tilde{c}_{i,\alpha,\sigma} &=& c_{i,\alpha,\sigma} - \bar{c}_{i,\alpha,\sigma},
\eea
where $\bar{c}_{i,\alpha,\sigma}$ and $\tilde{c}_{i,\alpha,\sigma}$ denote the projected annihilation operators in the flatband and the complementary Hilbert-Fock space, respectively. It is important to note that the projected operators satisfy the following anticommutation relations:
\begin{align}
\label{commutation}
\{\bar{c}_{i,\alpha,\sigma}, \bar{c}_{j,\beta,\sigma'}^\dagger \} &= \delta_{\sigma\sigma'}\, P_{\alpha\beta}(i-j), \nonumber \\
\{\tilde{c}_{i,\alpha,\sigma}, \tilde{c}_{j,\beta,\sigma'}^\dagger \} &= \delta_{\sigma\sigma'}\, \bigr(\delta_{\alpha\beta} - P_{\alpha\beta}(i-j)\bigl), \nonumber \\
\{\bar{c}_{i,\alpha,\sigma}, \tilde{c}_{j,\beta,\sigma'}^\dagger \} &= 0.
\end{align}

We can define the $\bar{\eta}$-pairing in the projected fermion operators in the flatband subspace as, 
\bea
    \bar{\eta} = \sum_\mathbf{r} \bar{c}_{\mathbf{r},\uparrow} \bar{c}_{\mathbf{r},\downarrow}.
\eea
However, in multi-band systems, the $\bar{\eta}$-pairing does not possess the SU(2) symmetry since
\begin{align}
    [H_{\textrm{total}},\bar{\eta}^\dagger]&=(U-2\mu)\bar\eta^\dagger \notag \\ &-U\Bigl[\sum_{j,\beta}\sum_{\substack{m\neq n \\ \gamma\neq\delta}}P_{\gamma\beta}(j-m)c_{m,\gamma,\uparrow}^\dagger P_{\delta\beta}(j-n)c_{n,\delta,\downarrow}^\dagger \notag \\ 
    &+ \sum_{i,j,\alpha,\beta}P_{\alpha\beta}(i-j)\,n_{i,\alpha,\uparrow}\bar c_{j,\beta,\uparrow}^\dagger c_{i,\alpha,\downarrow}^\dagger \notag \\
    &+ \sum_{i,j,\alpha,\beta}P_{\alpha\beta}(i-j)\, n_{i,\alpha,\downarrow} c_{i,\alpha,\uparrow}^\dagger \bar c_{j,\beta,\downarrow}^\dagger\Bigr].
\end{align}
 It has been shown that one can define a generalized $\eta$-pairing in a certain types of lattice systems including bipartite systems~\cite{PhysRevB.102.085140}. However, as we show below, the general flatband system does not fall in this case.

\subsection{Schriffer-Wolff transformation of $\eta$-pairing}

As a concrete example, we consider the one-dimensional Creutz ladder model that hosts a flatband (See Fig. \ref{fig:Creutz}(a)) with the onsite Hubbard interactions as, 
\bea
&H&_\mathrm{kin} = \sum_{i,\sigma} 
\Bigl[t\, c_{i+1,A,\sigma}^\dagger c_{i,A,\sigma} + c_{i+1,B,\sigma}^\dagger c_{i,A,\sigma}
\\
&+&c_{i+1,A,\sigma}^\dagger c_{i,B,\sigma} + c_{i+1,B,\sigma}^\dagger c_{i,B,\sigma}
+ t'\, c_{i,B,\sigma}^\dagger c_{i,A,\sigma} + \mathrm{h.c.}\Bigr], \notag \\
&H&_\mathrm{int} = U \sum_{i,\alpha=A,B} n_{i,\alpha,\uparrow}\,n_{i,\alpha,\downarrow},
\eea
where $\alpha$ and $i$ deontes sublattice and site index respectively. The kinetic part of the Hamiltonian consists of intra-cell hopping  $t'$  between sublattices A and B within the same unit cell, and inter-cell hopping $t$ between nearest unit cells.

Without the Hubbard interaction, the Hamiltonian possesses the eigenstates of the flatband, given as $U(\mathbf{k}) =  (1,-1)^\text{T}/\sqrt{2}$ . The flatband eigenstate renders the projection operator $P_{\alpha\beta}$ momentum independent. Consequently, the local projected fermion annihiliation operator can be written as,
\bea
\bar{c}_{i,\alpha,\sigma}  
= \frac{1}{2}\Bigl(c_{i,A,\sigma} - c_{i,B,\sigma}\Bigr),
\eea
which simplifies to a local linear combination of the original operators. in other words, $\bar{c}_{i,\alpha,\sigma}$ corresponds to the compact localized state (CLS) of the Creutz ladder model.

We derive the effective Hamiltonian in the flatband subspace using SW transformation by treating for the effect of $H_\text{int}$ perturbatively~\cite{BRAVYI20112793}. We define the many-body projection operators $\hat{\mathcal{P}}$ onto the flatband space and  its orthogonal projector $\hat{\mathcal{Q}} = \mathbf{1}-\hat{\mathcal{P}}$ onto the complementary space such that the fock state $|\psi\rangle=\gamma^\dagger_{n_1,\mathbf{k}_1,s_1}\gamma^\dagger_{n_2,\mathbf{k}_2,s_2}...\gamma^\dagger_{n_{n_f},\mathbf{k}_{n_f},s_{n_f}}|0\rangle$ satisfy $\hat{\mathcal{P}}|\psi\rangle=|\psi\rangle$ if $n_1=n_2=...=n_f=1$ otherwise zero. The effective Hamiltonian can be perturbatively expanded as,
\begin{align}
    \label{SWHamiltonian}
    H_\mathrm{eff} = (H_\mathrm{kin}-\mu N)\hat{\mathcal{P}} + \hat{\mathcal{P}}H_\mathrm{int}\hat{\mathcal{P}} &+ \frac{1}{2}\hat{\mathcal{P}}[\mathcal{L}(H_\mathrm{int}),\mathcal{O}(H_\mathrm{int})]\hat{\mathcal{P}}\notag \\
 &+\cdots,
\end{align}

where the superoperators $\mathcal{O}(X)$ and $\mathcal{L}(X)$ acting on an operator $X$ are defined as,
\[
\mathcal{O}(X) = \hat{\mathcal{P}}X\hat{\mathcal{Q}} + \hat{\mathcal{Q}}X\hat{\mathcal{P}}, \quad
\mathcal{L}(X) = \sum_{i,j} \frac{\ket{i}\bra{i}\,\mathcal{O}(X)\,\ket{j}\bra{j}}{E_i - E_j},
\]
 where $\ket{i}$ and $\ket{j}$ are eigenstates of unperturbed Hamiltonian $H_\text{kin}$. $E_i$ and $E_j$ are the corresponding eigenenergies. The operator $\mathcal{O}(X)$ extracts the off-diagonal components of $X$ that connect states in flatband and complementary subspaces, ensuring that the indices $i$ and $j$ in the summation belong to different subspaces. This guarantees that the energy denominator $E_i - E_j$ is always nonzero, as long as the energy gap between the flatband and the dispersive band exist.
 Physically, $\mathcal{L}(X)$ incorporates the effects of virtual tunneling process between the flat band and the complementary bands, with each contribution weighted by the inverse energy difference, thereby capturing the higher-order corrections to the effective low-energy dynamics.

For perfectly isolated flatband (i.e. infinite energy gap between the flatband and dispersive band), only the first two terms on the right-hand side of Eq. \eqref{SWHamiltonian} are non-vanishing such that
\begin{align}
    H_\mathrm{eff} &\approx(H_\mathrm{kin}-\mu N)\hat{\mathcal{P}} + \hat{\mathcal{P}}H_\mathrm{int}\hat{\mathcal{P}} \notag\\
    &= (H_\mathrm{kin}-\mu N)\hat{\mathcal{P}} + U\sum_{i\alpha} \bar c_{i,\alpha,\uparrow}^\dagger \,\bar c_{i,\alpha,\downarrow}^\dagger\bar c_{i,\alpha,\downarrow} \,\bar c_{i,\alpha,\uparrow}\hat{\mathcal{P}} \notag \\ 
    &=  (H_\mathrm{kin}-\mu N)\hat{\mathcal{P}} + U\sum_{i\alpha} \bar{n}_{i,\alpha,\downarrow}\bar{n}_{i,\alpha,\uparrow}\hat{\mathcal{P}},
\end{align}
where $\bar{n}_{i\alpha\sigma} = \bar c_{i,\alpha,\sigma}^\dagger\bar c_{i,\alpha,\sigma}$. Since $(H_\mathrm{kin}-\mu N)\hat{\mathcal{P}}$ is a constant times number operator, we can redefine the term with effective chemical potential as $(H_\mathrm{kin}-\mu N)\hat{\mathcal{P}} = \mu_\text{eff}N$. Then, from Eq. \eqref{commutation}, one can  verify that the effective Hamiltonian preserves the SU(2) symmetry, and the projected $\eta$ operator satify the SGA as,
\bea
\label{FlatSGA}
[H_\text{eff},\bar{\eta}^\dagger] = \left(\frac{U}{2}-2\mu_\text{eff}\right)\bar\eta^\dagger,
\eea
with
\bea
\bar \eta^\dagger
= \sum_{i,\alpha} \bar c_{i,\alpha,\uparrow}^\dagger \,\bar c_{i,\alpha,\downarrow}^\dagger.
\eea 
 
 In contrast, when the flatband is not perfectly isolated, second-order and higher-order terms from the SW transformation become significant. These additional corrections modify the effective Hamiltonian and disrupt the exact commutation relations of the SGA, so that $\bar\eta^\dagger$ no longer generates eigenstates with constant energy shifts.

\section{Modified spectrum generating algebra}
\label{Sec.Symmetry Breaking and Spectrum-Generating Algebra}

\begin{figure}[t]
    \centering
    \includegraphics[width=\linewidth]{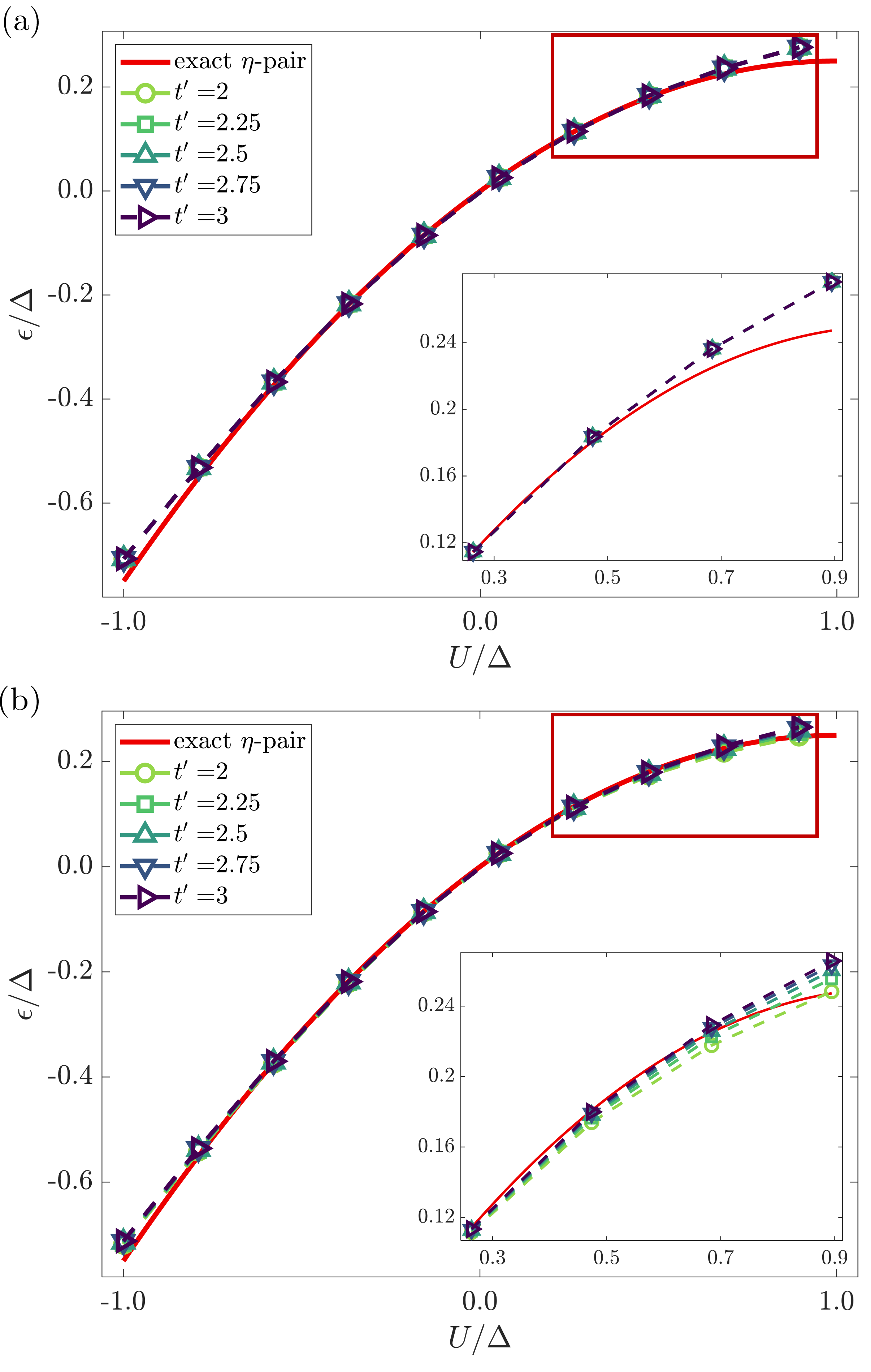}
    \caption{(a) Energy shift from the vacuum $\epsilon/ \Delta$ as a function of the scaled interaction strength $U / \Delta$ for various values of $t'$ when $t = 0$. Numerical simulations (symbols with dashed lines) agree closely with the analytical prediction $\epsilon = \frac{U}{2} - \frac{U^2}{4\Delta}$ (solid line) from the SW transformation, indicating the validity of the SGA in $t\ll t'$ limit. The inset corresponds to the magnified region highlighted by the red box in the main figure. (b) Energy shift from the vacuum $\epsilon / \Delta$ for $t = 0.5$, showing deviations from the analytical prediction due to enhanced interband coupling as $t'$ gets smaller. The results highlight the breakdown of the SGA as $t$ increases, emphasizing the impact of higher-order corrections in the effective Hamiltonian.
}
    \label{fig:SGA}
\end{figure}
In the perfectly isolated flatband limit the operator $\bar{\eta}^\dagger$ commutes with the Hamiltonian, generating an equally spaced tower of $\eta$-paired eigenstates. In the case of a single $\eta$ pair, the modified $\eta$-pairing state converges to the exact $\eta$-pairing state in the limit of an infinite band gap ($\Delta\to\infty$). However, with finite band gaps between the flatband and the dispersive band, interband coupling introduces nontrivial corrections that break the exact pseudo-spin SU(2) symmetry, thereby modifying the idealized spectrum generating algebra.

To capture these corrections, virtual hopping processes between the flat and dispersive bands are systematically incorporated using the Schrieffer–Wolff transformation. (See Fig. \ref{fig:Creutz}(b)) In the limit where the dispersive band remains nearly flat($t \ll t'$ limit), the band gap $\Delta$ can be treated as approximately constant, allowing for a perturbative expansion. In this regime, the second-order correction in Eq. \eqref{SWHamiltonian} can be explicitly computed as follows,
\begin{align}
\nonumber
    H_\mathrm{SW}^{(2)}&=- \frac{\hat{\mathcal{P}}H_\mathrm{int}\hat{\mathcal{Q}}H_\mathrm{int}\hat{\mathcal{P}}}{\Delta}
\\&=
\nonumber
-\frac{U^2}{\Delta} \sum_{i,j,\alpha,\beta} \hat{\mathcal{P}} \bar{c}_{i\alpha\uparrow}^\dagger\bar{c}_{i\alpha\downarrow}^\dagger\tilde{c}_{i\alpha\downarrow}\tilde{c}_{i\alpha\uparrow}\tilde{c}_{j\beta\uparrow}^\dagger\tilde{c}_{j\beta\downarrow}^\dagger\bar{c}_{j\beta\downarrow}\bar{c}_{j\beta\uparrow}\hat{\mathcal{P}}  
    \\&= -\frac{U^2}{4\Delta} \sum_{i,\alpha} \bar{n}_{i,\alpha,\downarrow}\bar{n}_{i,\alpha,\uparrow}\hat{\mathcal{P}}
\end{align}

The second-order correction to the effective Hamiltonian introduces an additional interaction term, which effectively renormalizes the on-site interaction strength to the $U_\textrm{eff}=U-\frac{U^2}{2\Delta}$. The virtual process contributes to the additional correction to the SGA, which is given as, 
\bea
\label{FlatSGA}
[H_\text{eff},\bar{\eta}^\dagger] = (\frac{U_\textrm{eff}}{2}-2\mu)\bar\eta^\dagger,
\eea

Despite the breakdown of exact SU(2) symmetry, the modified SGA in the perturbative regimes still indicate the logarithmic EE. As depicted in Fig.\ref{fig:SGA} (a) as $U$ becomes larger, higher order corrections become significant, energy shift deviate from the calculated $\epsilon$. Similarly in Fig.\ref{fig:SGA} (b), the deviation become larger as the band gap $\Delta$ gets smaller. Nonetheless, for $|U| \ll \Delta$, the structure of SGA remains robust enough to produce towers of equal level spacing. As the band gap $\Delta$ decreases (or as $U$ grows larger), the higher-order corrections in SW transformation become non-negligible, resulting in modification of the overall thermalization behavior. We find that the deviations of the equal level spacings.

\begin{figure}[t]
    \centering
    \includegraphics[width=\linewidth]{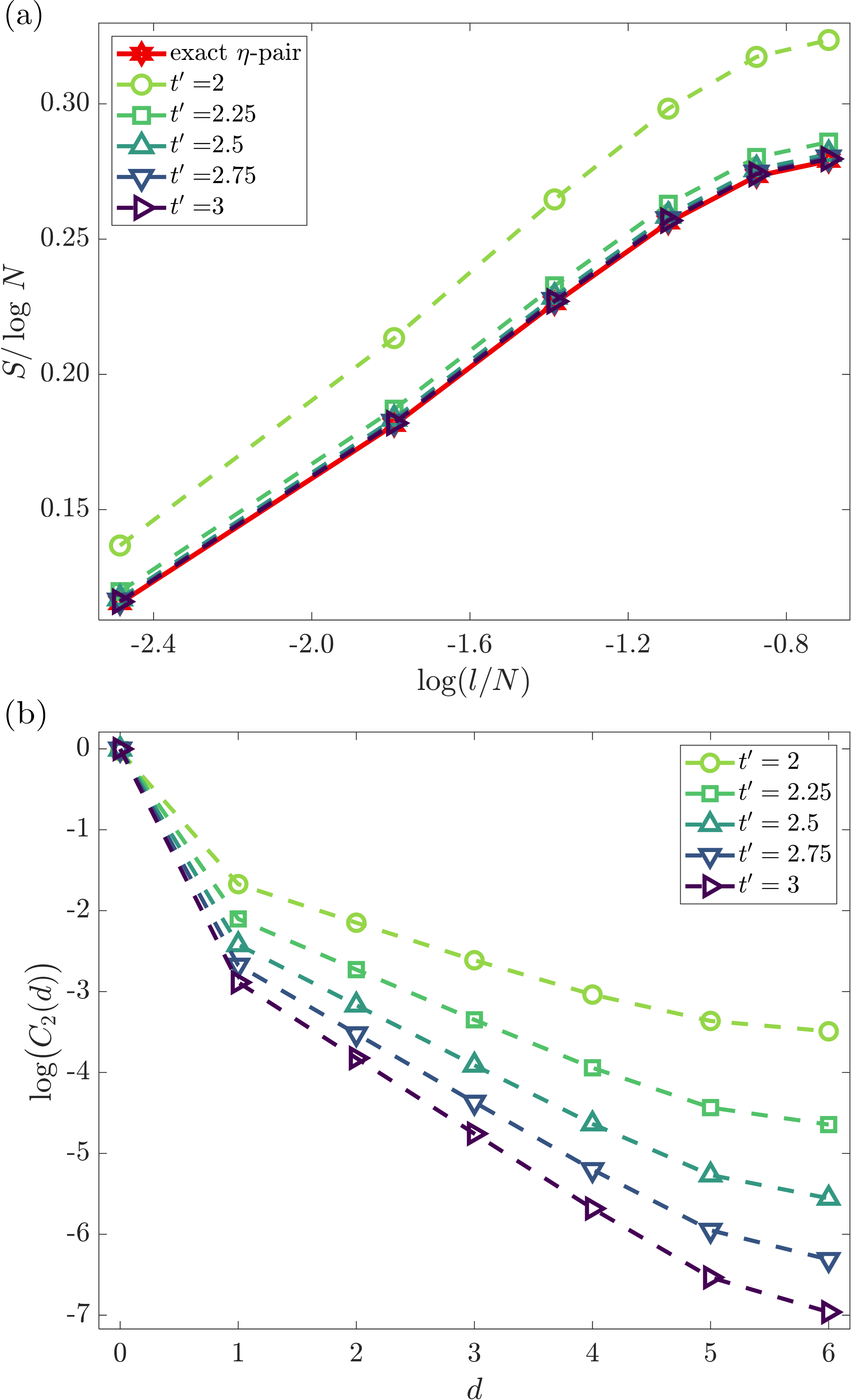}
    \caption{(a) Entanglement entropy scaling for the modified $\eta$-pairing states as a function of subsystem size $l$ (on a log scale) for a system of 12 unit cells and single $\eta$ pair, with $t=1$, $U=-1$, and various values of $t'$. The red solid line represents the entanglement entropy of the exact $\eta$-pairing state. Even in the band-touching limit ($t'=2$), the modified $\eta$-pairing state displays a logarithmic EE scaling. (b) Logarithm of two-point correlation function $C_2(d)$ in the same system configuration with Fig. \ref{fig:fig3}(a). Here, $d$ is the distance the unit cells in PBC hosting the up and down spins of a doublon. The linear tail of the $\text{log}\bigl(C_2(d)\bigr)$ illustrates broadening of the doublon wave function and its exponentially decaying tail. Steepening of the $\text{log}\bigl(C_2(d)\bigr)$ at larger $t'$, indicates localization of the doublon wave function, regaining spatial confinement of the exact $\eta$-pairing states.}
    \label{fig:fig3}
    
\end{figure}

\section{Entanglement entropy logarithm}
\label{Sec.Modified eta-Pairing States}

Our numerical results reveal that the entanglement entropy (EE) of the modified $\eta$-pairing state exhibits an approximately logarithmic scaling with subsystem size $l$, even in the band-touching limit ($t'=2$).  Fig~\ref{fig:fig3}(a) shows the calculated EE as a function of subsystem size $l$ for various values of the intra-cell hopping $t'$. We find that the modified state follows the logarithmic behavior approaching to the exact $\eta$-pairing state (red line) although the deviations from the exact $\eta$-pairing become apparent when interband coupling increases.

The modifications of the EE can be heuristically understood by the spatial structure of the pairing state. In the isolated flatband limit, the $\eta$-pair is strictly confined to a single CLS: 
\bea
\eta^\dagger\approx\frac{1}{2}\sum_i (c^\dagger_{i,A,\uparrow}-c^\dagger_{i,B,\uparrow})(c^\dagger_{i,A,\downarrow}-c^\dagger_{i,B,\downarrow}).
\eea 
When the virtual tunneling process is partially accessed via interband coupling, the pairing amplitude begins to extend onto neighboring sites, leading to the delocalization of the doublon wavefunction. 

This effect is captured by the calculations two-point correlation function, which is given as,
\bea
C_2(d) = \sum_{\substack{i,j \\ d(i,j)=d}} \sum_{\alpha,\beta} \left\langle c_{i,\alpha,\uparrow}^\dagger\, c_{j,\beta,\downarrow}^\dagger\, c_{i,\alpha,\uparrow}\, c_{j,\beta,\downarrow} \right\rangle,
\eea
 where the distance $d(i,j)$ between two unit cells $i$ and $j$ in PBC is defined by $d(i,j) = \min\{|i-j|,\,L-|i-j|\},$ with $L$ being the total number of unitcells in the system. Physically, $C_2(d)$ corresponds to the probability of observing the configuration of the up spin and down spin separated by distant $d$. Fig.~\ref{fig:fig3}(b) shows the calculated correlation function that shows the exponential decay as a function of the spatial separation of the doublon pair $d$. Importantly, in the large band gap ($t'$) limit the two-point correlation function becomes sharp again, recovering the spatially confined doublon characteristic of the exact $\eta$-pairing state. The exponential confinement of the doublon pair implies the robust EE logarithmic behavior. When $t'$ approaches to $2t+U$, the correlation function shows the Friedel oscillations signifying the failure of the EE logarithm.

\section{Doublon-Doublon interactions in many $\eta$-Pairing States }
\label{Sec.MultiEta}

The $\eta$-pairing state converges to the doublons of the exact CLS states in the isolated flatband limit ($\Delta\to\infty$). Due to the exact confinement of the CLS state, the eta pair states does not interact each other.

\begin{figure}[H]
    \centering
    \includegraphics[width=\linewidth]{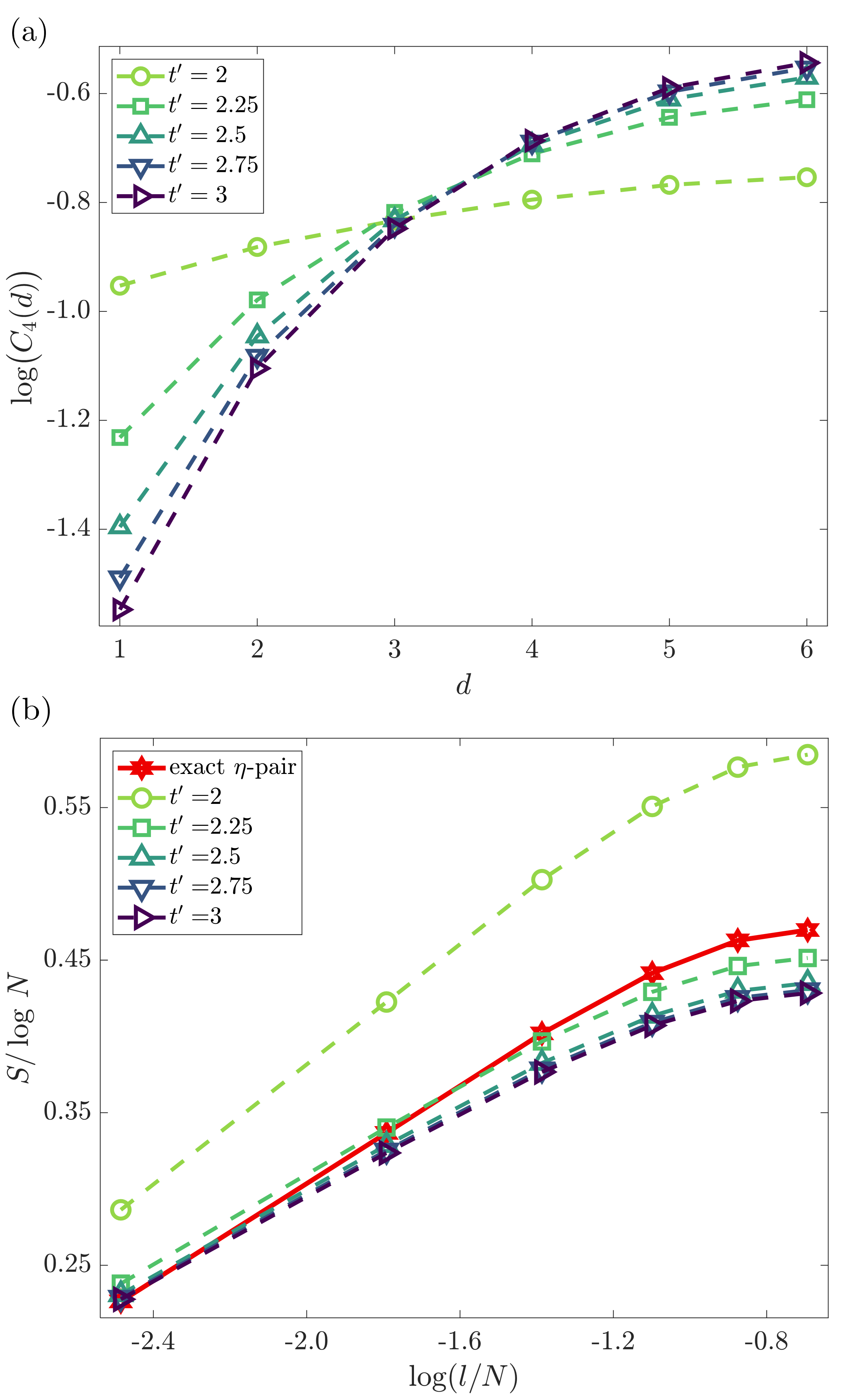}
    \caption{(a) Logarithm of four-point correlation function $C_4(d)$ in case of two $\eta$ pairs for a system of 13 unit cells, with $t=1$, $U=-1$, and various values of $t'$. $d$ is the distance between unit cells occupied by localized doublons in PBC. It can be seen from increasing of $C_4(d)$ that the state exhibiting distant doublon pair is more preferred due to the repulsive interaction between doublons. This tendency become more pronounced as band gap ($t'$) become larger.(b) Entanglement entropy scaling for the modified $\eta$-pairing states in the case of three $\eta$ pairs as a function of subsystem size $l$ (on a logarithmic scale) for a system of 12 unit cells, with $t=1$, $U=-1$, and various values of $t'$. The red solid line represents the entanglement entropy of the exact multi $\eta$-pairing state in the isolated flatband case. Although the modified $\eta$-pairing states converge in the large band gap limit, they do not coincide with the exact $\eta$-pairing states. }
    \label{fig:MultiEtaEE}
\end{figure}

However for any finite $\Delta$, the tails of the extended doublon wave functions overlaps. As a result, the accumulations of the multiple $\eta$ pairs undergoes inter $\eta$-pairing interactions. The dominant two-body interaction between $\eta$-pair is repulsive regardless of the sign of $U$ due to the Pauli exclusion principle. This feature can be calculated from the diagonal matrix element of four-point correlation function $C_4$, which is given as,

\bea
C_4(d)=\sum_{\substack{i,j \\ d(i,j)=d}} \langle D_{i,j}^\dagger D_{i,j}\rangle 
\eea
where $D_{i,j}^\dagger = \sum\limits_{\alpha,\beta,\gamma,\delta} c_{i,\alpha,\uparrow}^\dagger c_{i,\beta,\downarrow}^\dagger c_{j,\gamma,\uparrow}^\dagger c_{j,\delta,\downarrow}^\dagger$ is doublon pair creation operator at $i$ and $j$ unit cells. In physical terms, $C_4(d)$ represents the probability of observing a configuration where doublons are separated by distant $d$. Fig. \ref{fig:MultiEtaEE}(a) illustrates behavior of the distance-dependent four-point correlation function $C_4(d)$ for a system of 13 unit cells, with $t=1$ and $U=-1$. The result shows that $C_4(d)$ increases as doublons become more widely separated, reflecting a preference for distant pair configurations. This behavior arises from the repulsive interaction between $\eta$-pairs. Moreover, as the band gap $t'$ grows, this tendency becomes more significant. 

The preservation of doublon pairs under repulsive interactions can lead to an intriguing scenario where, despite the inherent repulsion, the system's ground state transitions into a Bose–Einstein condensate (BEC). The doublon pairs act as composite bosons, remaining robust even in the presence of strong repulsive forces. This stability allows the pairs to condense into a single macroscopic quantum state at low temperatures, manifesting as off-diagonal long-range order characteristic of a BEC.  When $U$ is negative, the $\eta$-pairing state becomes the ground state in each fixed particle-number sector. In this regime, the ground state of the system behaves as composite bosons like Cooper pairs, leading to a superconducting ground state with off-diagonal long-range order. Within a simple mean-field picture, the critical temperature $T_c$ is expected to scale roughly with $|U|$, although a detailed prediction would require further analysis of the band structure and interband coupling effects.

Despite this deviation, key hallmarks of $\eta$-pairing remain robust. In particular, the EE of the modified multi‑$\eta$ pairing states is even lower than that of exact $\eta$-pairing states and follows a logarithmic scaling with the subsystem size. Fig.~\ref{fig:MultiEtaEE}(b) (which is analogous to Fig.~\ref{fig:fig3}(a) but for the three-$\eta$ pair case) clearly shows that, even when the exact tower structure is lost, the EE retains a log‑law behavior with even lower value. This indicates that the underlying pairing order remains strong despite the breakdown of the exact pseudo-spin SU(2) symmetry.

\section{Discussion}
\label{Sec.discussion}

In this work we have investigated the deformation of $ \eta $-pairing states in flatband systems due to interband coupling, using the Creutz ladder as an example. Our analysis, demonstrates that even when the flatband is not perfectly isolated, a subset of eigenstates retains key characteristics of $ \eta $-pairing. In the single‑$ \eta $ pair regime, the modified $ \eta $-pairing state converges to the exact $ \eta $-pairing state only in the large band gap limit ($\Delta\to\infty$); for any finite gap, SW transformation shows that virtual excitations into the dispersive band produce shift that deviates from the ideal value as the interaction strength increases or the gap decreases. This quantifies the effect of interband coupling to the pseudo-spin SU(2) symmetry.

We have characterized these modified states through two key observables. First, numerical simulations that the entanglement entropy of the modified $ \eta $-pairing state exhibits an approximately logarithmic scaling with subsystem size $l$, even in the band-touching limit, in contrast to the generic eigenstates. Second, the real-space two-point correlation function shows that while the exact $ \eta $-pairing state features a sharply localized doublon confined to a CLS, the modified state exhibits a broadened doublon profile due to interband hybridization. Notably, this broadening is reduced in the large band gap limit, thereby restoring the sharp localization characteristic of the isolated flatband.

Despite the breakdown of the exact pseudo-spin SU(2) symmetry and the accompanying disruption of the SGA, the many‑$ \eta $ states still exhibit remarkably low entanglement entropy that follows logarithmic scaling. This persistence of low entanglement entropy, alongside spatially localized pairing correlations, underscores the robustness of the underlying ODLRO even when the ideal eigenstate structure is lost. 

Collectively, our findings shed light on the interplay between interaction-driven pairing and band structure effects in realistic flatband systems. The persistence of $ \eta $-pairing signatures, such as low entanglement entropy and confined doublon correlations, even in the presence of interband coupling, suggests that strict band isolation is not an absolute requirement for robust pairing phenomena. This has significant implications for experimental platforms such as ultracold atoms, photonic lattices, and designer electronic systems, where perfect flatband isolation is often unattainable.

Optical lattices in cold-atom systems offer a controllable platform for realizing the flat-band structures. There has been experimental realizations of the Lieb lattices in both bosonic and fermionic platforms by tuning the lattice unitcell ~\cite{doi:10.1126/sciadv.1500854,lebrat2024ferrimagnetismultracoldfermionsmultiband}. The control of the lattice distance can effectively tune the intra and intercell hopping ratio $t'/t$ in addition to the interaction strength $U$. The wide range of the  controllability of the interaction strength ($U/t\approx 1 \sim 10$ via Feshbach resonance) can explicitly measures the dynamics of the many-body phase in the controllable manner~\cite{RevModPhys.82.1225,cite-key,doi:10.1126/science.aaa7432,doi:10.1126/science.aaf8834}. In such setups, $\eta$-pairing states can be prepared via adiabatic ramping process starting from a configuration with localized doublons~\cite{PhysRevLett.104.240406}. Their nonthermal behavior can be signified by a persistently high probability of spin–spin correlations, seen as a sharp peak in the pair momentum distribution~\cite{PhysRevLett.104.240406,PhysRevLett.123.030603}.

Future work may extend our analysis to the dynamical response of these pairing states and the exploration of similar phenomena in higher-dimensional flatband systems (such as Lieb and Kagome lattice). While the spatial dimensions of the system may results in the qualitative changes in the dynamical behavior, it is important to note that the SW transformation does not alter the results. In our case the Creutz lattice model exhibit the lower energy flatband with quadratic band touching. This energy band coincides with the Lieb lattice.

\section{Acknowledgement} 
M.J.P. thanks the helpful discussions from Kyoung-Min Kim and Beom Hyun Kim.
This work was supported by the National Research Foundation of Korea
(NRF) grant funded by the Korea government (MSIT) (Grants No. RS-2023-00218998). This work was supported by the BK21 FOUR (Fostering Outstanding Universities for Research) program through the National Research Foundation (NRF) funded by the Ministry of Education of Korea.

\bibliography{eta}
\end{document}